\documentclass[article,amsart,fleqn,amsmath,floats,nofootinbib,aps,twocolumn,pra,floatfix,showpacs,superscriptaddress]{revtex4-2}

\usepackage{cmap}
\usepackage{wrapfig} 
\usepackage[pdftex]{graphicx}
\usepackage{pgfplotstable}
\usepackage[pdftex,unicode,colorlinks,bookmarks=false,citecolor=blue,linkcolor=blue,urlcolor=blue]{hyperref} 
\usepackage[T1,T2A]{fontenc} 
\usepackage{latexsym}
\usepackage{amsmath}
\usepackage{amssymb}
\usepackage{amsfonts}
\usepackage{color}
\usepackage{bm}
\usepackage{verbatim}
\usepackage[english]{babel}
\usepackage[utf8]{inputenc}
\usepackage{euscript}
\usepackage{subfigure}
\usepackage[normalem]{ulem}
\usepackage{listings}
\hypersetup{
colorlinks=true,
linkcolor=blue,
filecolor=magenta,
urlcolor=cyan,
}

\begin{document}
\title{Reconstruction of the Electron–Phonon Interaction Function for Superconductors Using Inhomogeneous Point-Contacts and Background Correction in Yanson Spectra}
\author{N.~L.~Bobrov}
\affiliation{B.~Verkin Institute for Low Temperature Physics and Engineering, National Academy of Sciences of Ukraine\\
prosp. Nauki 47, 61103 Kharkov, Ukraine}

\email{bobrov@ilt.kharkov.ua}

\published {\href{https://doi.org/10.31857/S0044451021070087}{\emph{JETPr}, \textbf{160}(1), 73, (2021)};
[\href{}{\emph{JETPe}, \textbf{133}(1), ??, (2021)}]}
\date{\today}

\begin{abstract}\emph{The application of inhomogeneous niobium point-contacts in the superconducting state for reconstructing the electron–phonon interaction function is considered. The method is based on the use of the non­linearity of current–voltage curve, which is due to the inelastic processes of suppressing excess point-contact current when nonequilibrium phonons are scattered from the electrons undergoing Andreev reflection. A new model of background occurrence in point-contact Yanson spectra and some ways to correct this background are proposed}.


\end{abstract}

\maketitle

\tableofcontents{}

\maketitle

\section{INTRODUCTION}
The spectroscopy of electron–phonon interaction (EPI) of point-contacts with direct conductivity is based on duplication of carriers: in the current-conducting state electrons are divided into two groups, for which the energy difference between the occupied and free electrons states on the Fermi surface is several millielectronvolts, i.e., is equivalent to the applied voltage \cite{1,2}. The number of studies devoted to the determination of the EPI function using point-contacts in the normal state exceeds several hundred; two monographs generalizing the results obtained in this field have been published \cite{3,4}.

For the superconducting state, the use of the nonlinearities of the excess current of ballistic point contacts in the phonon energy range for reconstructing the EPI function was justified theoretically in the studies by Khlus and Omel’yanchuk in 1983 \cite{5,6}. However, their results found experimental confirmation only almost 30 years after \cite{7,8}. This time delay was caused by the existence of superconducting background (disregarded by the theory), which led to the necessity of using the difference in the second derivatives of current–voltage ($I–V$) curves in the superconducting and normal states for this procedure, whereas the theory predicted that the EPI function is proportional to the first derivative of excess current or the difference in the first derivatives for the superconducting and normal states. The mechanism of the formation of excess-current nonlinearity in the phonon energy range is related to the scattering of nonequilibrium phonons, generated by electrons with excess energy (equivalent to the voltage applied to the contact), from the Andreev electrons. The nonlinearity of the normal-state $I–V$ curve is formed in a volume with a characteristic size of the same order of magnitude as the contact diameter. This is related to the geometric limitations of the backscattering processes responsible for this nonlinearity. The backscattering processes are considered to be the processes of electron scattering from nonequilibrium phonons, in which an electron returns as a result of scattering to the electrode it was emitted from.

The processes of scattering of nonequilibrium phonons from Andreev electrons are not restricted by these limitations and, in principle, may occur at any distance from the short-circuit in which they coexist. Nevertheless, the theory takes into account only the scattering processes occurring in the region of high current density, where the concentrations of both components are high. On the whole, the conclusions of the theory were confirmed for superconductors with a large coherence length, such as, e.g., tin or aluminum \cite{8}. At the same time, it was found for ballistic tantalum point-contacts that the region near a short­circuit with a characteristic size on the order of coherence length is also involved in the formation of excess­current nonlinearity \cite{7,9}. The contribution of this region turned out to be much larger than that from the region with a size on the order of contact diameter. The significant difference between the superconducting additive to the spectrum for tantalum and that for superconductors with a larger coherence length calls for application of peculiar procedures when reconstructing the EPI function using this additive. These procedures were found useful when studying the inhomogeneous superconducting point-contacts. The experimental data on the reconstruction of EPI functions for different superconductors from the additional nonlinearity arising as a result of the transition of point-contact with direct conductivity to the superconducting state were reviewed in \cite{10}.

For ballistic point-contacts the nonlinearity of the $I–V$ curve in the normal state is at least of the same order of magnitude or even larger than the additional excess-current nonlinearity, arising when a contact passes to the superconducting state. The necessity of selecting this nonlinearity in the pure form for reconstructing the EPI function reduces to a great extent the practical validity of this procedure, because the transfer of many superconductors to the normal state at low temperatures is highly hindered or even impossible for technical reasons.

One could get out of this dead-end situation if it were possible to cancel of the nonlinearity related to the backscattering processes responsible for the nonlinearity in the normal state, having retained at the same time the additional excess-current nonlinearity caused by the scattering of nonequilibrium phonons from Andreev electrons. Since this situation cannot be implemented in ballistic contacts, we will consider possible alternatives.

The spectrum intensity in the normal state decreases as a result of the contamination of material in the narrowing region, which is accompanied by the decrease in the elastic mean free path of electrons. A limiting case of this decrease is observed in diffusion point contacts, i.e., the contacts satisfying the condition ${{l}_{i}}\ll d\ll {{\Lambda }_{\varepsilon }}$, where $d$ is the contact diameter, $l_i$ is the elastic relaxation length, $l_\varepsilon$ is the energy relaxation diffusion length, ${{\Lambda }_{\varepsilon }}=\sqrt{{{l}_{i}}{{l}_{\varepsilon }}/3}$, and $l_\varepsilon$ is the energy relaxation length. The spectrum intensity in the diffusion mode is lower than that in the ballistic mode by a factor of $\sim l_{i}/d$ in order of magnitude.

In accordance with the theoretical model \cite{11}, the second derivative of the $I–V$ curve of diffusion contact differs from that for a ballistic contact in only lower intensity. In addition, due to the isotropization of electron-momentum distribution, one can observe almost complete isotropization of the EPI spectrum. This may lead to only a small broadening of spectrum and insignificant variation in its shape. The theory does not predict any other consequences of the reduction of electron elastic relaxation length.

As was mentioned above, the experimentally observed decrease in the elastic scattering length is due contamination of the point-contact, i.e., the increase in the concentration of impurities and lattice defects. Figures 1 and 2 from \cite{12} demonstrate an increase in the background and degradation of high-energy phonon peaks in copper point-contact spectra with an increase in the concentration of manganese and iron impurities. A method for direct determination of the electron elastic scattering length $l_i$ in a point-contact was reported in \cite{13}. For relatively small contacts, the transition from the ballistic to diffusion mode, when the elastic relaxation length becomes comparable with the characteristic lattice parameters, is accompanied by significant lattice distortions. Along with the diffusion of spectra, predicted by the theory, these distortions lead to strong suppression of high-energy phonon features, up to their complete disappearance, and to a high background level \cite{12,14}. Note that the suppression of high-energy phonons is also characteristic of the EPI functions reconstructed in tunnel experiments for films with a distorted lattice \cite{15}.

As was noted above, in view of the geometric limitations, backscattering processes are efficient in a contact volume with a characteristic size on the same order of magnitude as the contact diameter. However, for diffusion contacts under certain conditions, these processes can be efficient only near the contact center. If the maximum concentration of defects and impurities is obtained at the boundary between the electrodes and rapidly decreases at the contact periphery, the electrons scattered from nonequilibrium phonons will diffuse with high probability along the gradient of decreasing impurity concentration. In other words, the backscattering processes, determining the spectrum in the normal state, will be concentrated mainly near the electrode contact interface. This may lead to an increase in the surface-phonon contribution to the EPI spectrum and, correspondingly, to additional smearing of the spectrum. Obviously, if only one of the bank of the point contact is in the diffusion mode, then in the resulting spectrum the intensity of each partial contribution will be determined by the dirty bank.

Nevertheless, all phonon features remain at their positions in the normal state in these contacts with a very high background level, diffuse spectra, and strongly suppressed high-energy phonons \cite{12,14}. This is an unambiguous evidence for preservation of electron duplication in these contacts \cite{1,2}; i.e., they are spectroscopic.

Let us now return to consideration of the excess­current nonlinearity caused by the scattering of non­equilibrium phonons from Andreev electrons. As was noted above, in tantalum ballistic point contacts \cite{7,9}, the contact banks make a significant contribution to the additional nonlinearity of $I–V$ curve during the transition to the superconducting state because of the smaller coherence length. The formation of excess-current nonlinearity in these contacts occurs in a volume with a characteristic size on the order of coherence length. Obviously, the value of this nonlinearity in point-contacts from similar superconductors (with a comparable coherence length) will depend on both the excess current value and on the volume within which this nonlinearity is formed. Excess current for point contact in diffusion mode is 55\% of its value in ballistic contact, and the corresponding nonlinearity will be formed approximately in the same volume with a characteristic size on the order of coherence length; therefore, one might expect the additional nonlinearity of the $I–V$ curve of superconducting state in the diffusion mode to be only a half of that in the ballistic mode.

Thus, is appears that the transfer of contact to the diffusion mode may solve the above-stated problem: one has a significant decrease in the contribution of backscattering processes to the total nonlinearity and a relatively small decrease in the contribution of the scattering of nonequilibrium phonons from Andreev electrons. However, it is pertinent to remind here about the influence of defects and impurities on the shape of EPI spectra. As was noted in \cite{12,14}, the presence of strong lattice distortions leads to significant suppression of high-frequency modes and general blur of EPI spectra, which is confirmed by independent data of tunnel experiments \cite{15}. Since nonequilibrium phonons reflect the vibrational structure of material in the vicinity of their generation, to obtain the EPI spectrum of unperturbed material, the impurities and defects should be concentrated (in the ideal case) only near the contact center, and the energy relaxation diffusion length should exceed the size of the region within which the nonlinearity of $I–V$ curve is formed in the phonon frequency range in the superconducting state. Thus, the final solution of the above-stated problem is a contact with a nonuniform impurity distribution, consisting of a dirty nucleus (the point-contact region where backscatter­ing is efficient) and pure banks; hereinafter, this contact will be referred to as inhomogeneous. No additional limitations are imposed on the contact.

Let us outline at once the range of possible candidates applicable for studies with an inhomogeneous contacts. Superconductors with an extremely large coherence length can be rejected at once, because banks are not involved in the formation of nonlinearity in this case, and the volume in which lattice distortions are concentrated will be close to or coincide with the current concentration domain. For the same reason, study of high-temperature superconductors (HTSCs) is unlikely, but now because of the extremely small coherence length. All other superconductors lying in the intermediate range can be considered as possible candidates for testing the method.

Pressure point-contacts are the most appropriate objects for verifying the above suggestions, because surface is always more contaminated as compared with volume, and additional distortions are introduced at the inter electrode boundary during the point-contact formation.

\section{EXPERIMENTAL}
The object of investigation was the well-studied superconductor with a known EPI function: niobium. It is a fairly complex object from the point of view of both point-contact and tunnel spectroscopies. The oxides on its surface may vary in composition in wide limits and change their properties in dependence of the oxidation state. In addition, the oxide layer contains many depairing centers, which suppress superconductivity.

Point-contacts were formed using pure niobium with a resistance ratio of $\rho_{300}/\rho_{res} \sim100$, where $\rho_{300}$ and $\rho_{res}$ are, respectively, the resistivity at room temperature and the residual resistivity \cite{16,17}. The elastic relaxation length in niobium turned out to be $l_i\sim220$~nm, and the coherence length was $\xi_0 \sim44$~nm; with allowance for the elastic scattering, the reduced coherence length was $\zeta \sim36$~nm. Here, $\frac{1}{\zeta}=\frac{1}{\xi_0}+\frac{1}{l_i}$. Contacts were prepared according to the shear technique \cite{18,19}. Taking into account the importance of the role played by oxide on the electrode surface in providing the high quality and mechanical and electrical stability of point-contacts, two approaches were applied in the experiments. Native oxides were used in the first approach: before installing in the device for forming point-contacts, the electrodes were etched in a mixture of acids, washed, and then dried. In the second approach, electrodes processed in a similar way were placed in a sputtering plant and, after heating in vacuum to a premelting temperature, a thin aluminum layer was deposited on their surface, which was then oxidized in a boiling solution of hydrogen peroxide. The prepared electrodes were installed in the device for forming point­contacts \cite{20}. Hereinafter, the contacts formed between electrodes coated by native oxides will be referred to as first-type contacts and the contacts formed using the second approach as the second-type contacts.

\section{EXPERIMENTAL RESULTS AND THEIR PROCESSING}

\begin{figure}[]
\includegraphics[width=8.5cm,angle=0]{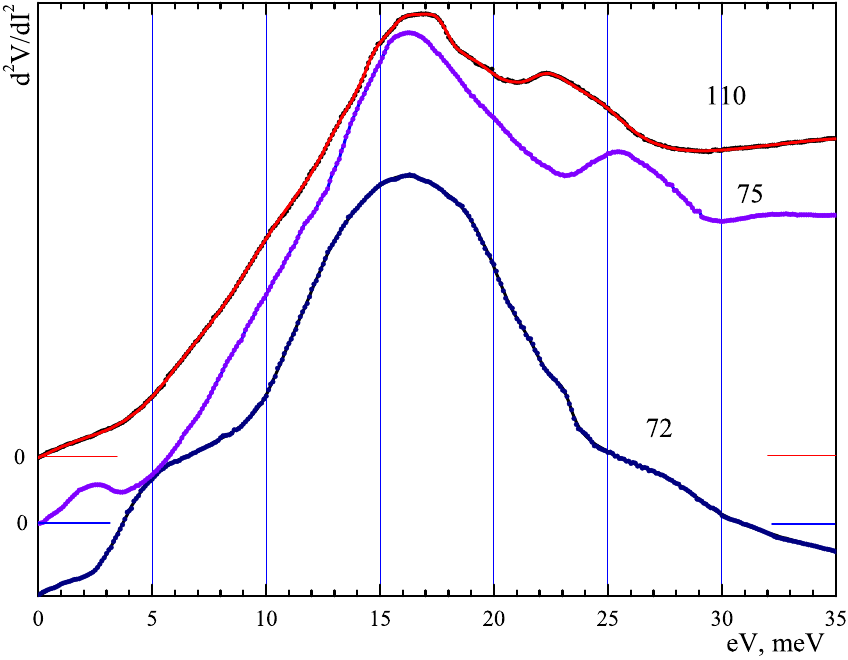}
\caption[]{EPI spectra for niobium point-contacts. Superconductivity is suppressed by magnetic field, $H\sim4\div5$~T, $T=4.2$~K. The experimental curves are smoothed by polynomial approximation.}
\label{Fig1}
\end{figure}

Despite the large statistics (more than a hundred of point-contacts of each type), even the best spectra in the normal state do not describe quite adequately the ballistic flight of electrons. The phonon features in them are significantly diffuse, the high-frequency phonons are strongly suppressed, and there is a high background level. In addition to the aforesaid, the high-frequency peak is blue-shifted in a number of spectra. Some spectra have a form of a wide diffuse maximum in the vicinity of the first peak, with a weak shoulder near the second peak. Typical representatives of these spectra are shown in Fig.\hyperref[Fig1]{1}. However, the overwhelming majority of the second derivatives of $I-V$ curves in the normal state for contacts of both types are curves of very low intensity without any pronounced phonon features. Unfortunately, the sparse contacts whose spectra in the normal state were similar to those presented in Fig.\hyperref[Fig1]{1} could not be transferred to the superconducting state, because only few contacts could withstand the magnetic field switch off.

At the same time, a complete set of characteristics could be recorded for some contacts from the much more numerous group, with spectra free of pronounced phonon features in the normal state, because of their much larger number. When these contacts passed to the superconducting state (at a constant level of modulating signal), the second-derivative intensity increased approximately by an order of magnitude, and the spectral shape changed radically.

\begin{figure*}[]
\includegraphics[width=16.5cm,angle=0]{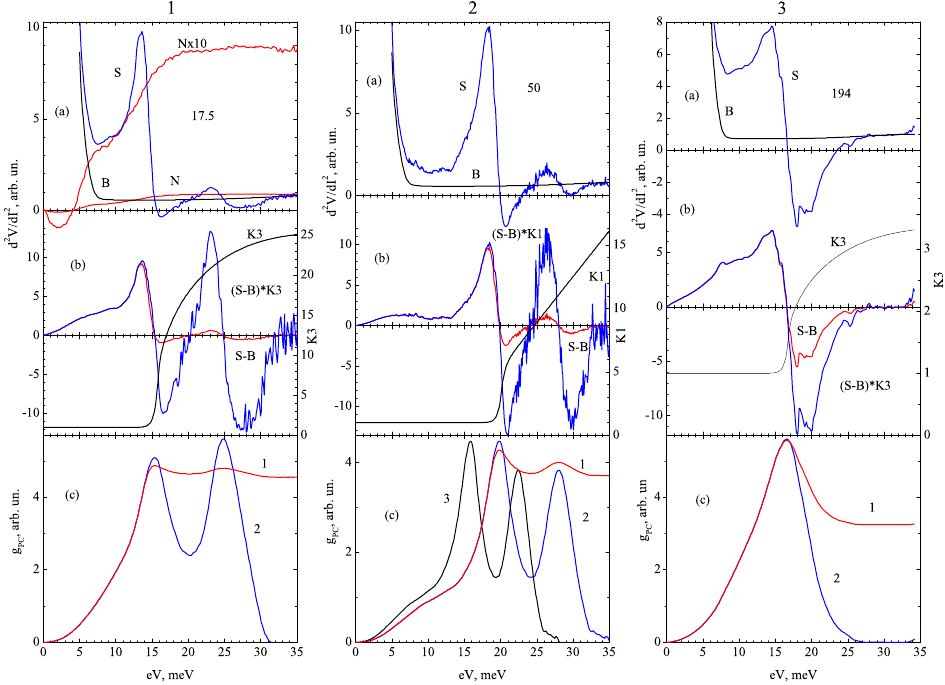}
\caption{Procedure of reconstruction of the EPI function for niobium using superconducting point-contacts: (a) $S$ are the second derivatives of $I–V$ curves of niobium point-contacts in the superconducting state and $B$ is the background curve; (b) $S–B$ is the second derivative with subtracted background, $(S–B)K$ is the second derivative with subtracted background after the correction, and $K$ is the correction curve; (c) point-contact EPI functions $g_{PC}$ (the abbreviation PC stands for point contact) obtained by integrating the second derivative of $I–V$ curve with subtracted background (panel (b)): (1) before the correction and (2) after the correction. For the 17.5~$\Omega$ contact: (a) $N$ is the second derivative of $I–V$ curve in the normal state, recorded with the same modulation current as the curve $S$; $N\times10$ is the same curve on the larger scale. For the 50~$\Omega$ contact: (c) curve 3 is obtained from curve 2 by multiplying the $x$ coordinates of points by 0.8 to compensate for the series resistance. Contacts 1 and 2 are of the second type, and contact 3 is of the first type.}
\label{Fig2}
\end{figure*}
Figure \hyperref[Fig2]{2.1a} shows an example of this transformation of the second derivative of $I–V$ curve for one of the second-type contacts. The shape of the second derivative in the superconducting state is similar to the difference between the second derivatives of $I–V$ curves in the superconducting and normal states of the tantalum point-contacts studied in \cite{9}. There are also some differences: the niobium spectrum does not exhibit the nonequilibrium feature and transformation of the soft phonon mode in spectrum from a shoulder in the normal state (at about 10~meV) into a peak when passing to the superconducting state. In tantalum point-contacts this transformation is due to the selection of phonons with small group velocities. The absence of this selection in niobium is explained by the much smaller (as compared with tantalum) volume in which nonequilibrium phonons interact efficiently with Andreev electrons and (or) insufficiently long elastic mean free path of nonequilibrium phonons. Tantalum is closer to the boundary of coherence-length range in comparison with other superconductors: the coherence length is sufficiently small to involve the near-contact region in the formation of superconducting-state nonlinearity and, at the same time, sufficiently large to implement efficient selection of phonons in this region.
\begin{figure*}[]
\includegraphics[width=16.5cm,angle=0]{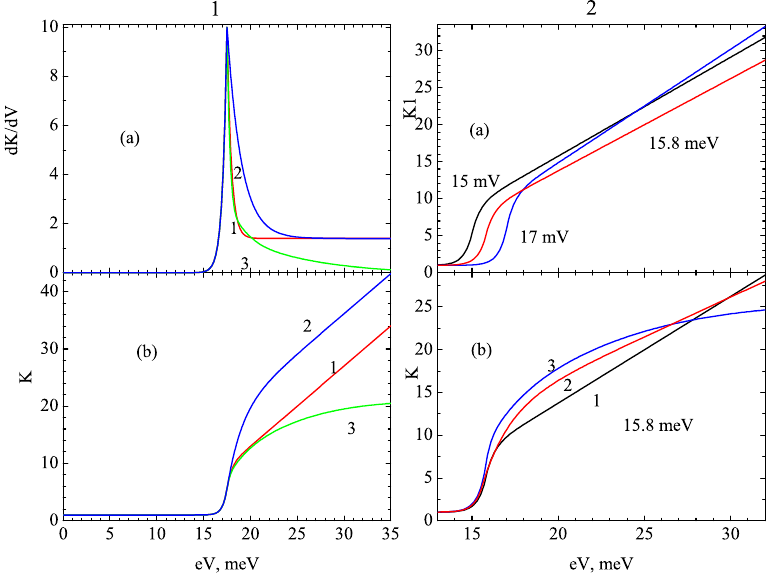}
\caption{1(a) Derivatives of the correction curves used as “preforms”. The ascending portions are identical for all preforms; they are segments of exponentials. The descending portions for preforms 1 and 2 are also segments of exponentials, while in "preform" 3 they consist of two segments of exponentials conjugated by a polynomial. 1(b) Correction curves 1-3, obtained from the "preforms" presented in panel 1(a). The position of the curves on the energy axis corresponds to the position of maximum in the preform. Using different correction curves, one can estimate the degree of influence of their forms and the position on the energy axis on the result of EPI function reconstruction.
2(a) Change in the scale of correction curve $K1$ in dependence of its position on the energy axis (numbers near the curves) for the point-contact shown in Fig.~\hyperref[Fig2]{2.1}. 2(b) Form of the correction curve at the points of magnitude minima (see formula\hyperref[eq__1]{(1)} and Fig.\hyperref[Fig4]{4}) for the point-contact in Fig.\hyperref[Fig2]{2.1}. The numbers near the curves indicate their position on the energy axis.}
\label{Fig3}
\end{figure*}
Despite the fact that the nonlinearity of $I–V$ curve in the normal state can be neglected for such contacts, one still must subtract the superconducting background from the second derivative of $I–V$ curve. When plotting the background curve, it is necessary to nullify the second derivative at voltages above the boundary of the phonon spectrum. At $eV\gg\Delta$, where $\Delta$ is the superconducting energy gap, the background is described by a smooth curve without any features, and the main requirement to the background curve near the gap feature is to exclude the occurrence of artifacts on the difference curve $S–B$. This also holds true for the missing portion (starting from zero) in the difference curve. In Figs.\hyperref[Fig2]{2.1a} and \hyperref[Fig2]{2.1b} one can see, respectively, the background curve satisfying these criteria and the difference curve. The missing portion for the difference curve at displacements from 0 to 5~mV is plotted by hand. Since the EPI function is identically zero beyond the phonon spectrum, to reconstruct correctly the EPI function, the difference curve must satisfy the geometric rule referred to as the \textbf{\emph{sum rule}}: \emph{the total areas under the integrable curve above and below the abscissa axis should be identical}.

Obviously, this rule is not satisfied for the curves presented in Fig.\hyperref[Fig2]{2}, because the high-energy part of the spectrum is significantly suppressed and must be corrected. As well as in the case of tantalum point contacts \cite{9}, the correction curve is unity in the low­energy range and exceeds unity in the region where the spectrum is suppressed. Multiplication of the difference curve by the correction curve yields a curve for which the sum rule is satisfied. The EPI functions reconstructed from the curves before and after the correction are shown in Fig.\hyperref[Fig2]{2c}.

The spectrum of the second-type contact (Fig.\hyperref[Fig2]{2.2}) is fairly interesting. Its shape is similar to that of the previous spectrum; however, all features in it are blue­shifted. After subtracting the background $B_S$, correction of the obtained curve, and subsequent integration, we obtain an EPI function with peaks located near 20 and 28~meV. This pattern corresponds to the EPI function of niobium, extended by 20\% along the x axis. Having multiplied the $x$ coordinates of the curve by 0.8, we obtain an EPI function fairly close to tunnel. The blue shift of the phonon features may be caused by the complex contact structure. Since the thickness of the deposited aluminum layer was monitored with insufficient accuracy, it likely exceeded the conventional value. During subsequent processing of electrodes in hydrogen peroxide the aluminum film was not oxidized throughout the entire thickness and played the role of a series resistance, making the phonon peaks blue-shifted; this shift can be taken into account if the true position of phonon features is known.

An inverted $S$-shaped curve without any additional structure (Fig.\hyperref[Fig2]{2.3}) is most typical of the first-type contacts. As well as for the previously considered contacts, the difference curve must be corrected after subtracting the background. The panel (c) shows the EPI functions reconstructed from the difference curve before and after the correction. Since the superconducting additive to the spectrum is formed in a volume with a characteristic size of the length of conversion of Andreev electrons into Cooper pairs, based on the shape of the EPI function reconstructed from this additive, one can draw an indirect conclusion about the degree of lattice perfection in this volume. As follows from Fig.\hyperref[Fig2]{2.3}, the EPI function reconstructed for this contact is a diffuse bell-shaped curve. This shape is characteristic of a highly deformed metal. A corresponding example can be observed in Fig. 1 in \cite{21} for the spectrum of a zirconium break junction. At the same time, a more sparing technique of shear pressure contacts provided zirconium spectra with a much better resolution, even with electrodes of obviously worse quality (see Fig.~9 in \cite{22}). Thus, the thickness of the defect layer in the first-type niobium contacts is most often comparable with the conversion length of Andreev electrons in Cooper pairs or even exceeds it. Therefore, when using inhomogeneous point-contacts for reconstructing the EPI function, the key condition is the minimally possible defect-layer thickness. It must be at least obviously smaller than the size of the region where the nonlinearity of I–V curve is formed in the superconducting state. This condition turned out to be much more easily satisfied for the second-type contacts.

Obviously, when plotting the background curve sticking to the above rules, possible variations in the same time, the influence of the shape of the correction curve and its position on the energy axis has been poorly studied. For tantalum point-contacts, as can be seen in Fig.~15 from \cite{9}, there is a spread in these parameters for different contacts. The correction curves were plotted in \cite{9} by hand, and their amplitude and position on the energy axis were chosen empirically. In this study, we attempted to reveal the general regularities for correction curves of different shapes.

\begin{figure}[]
\includegraphics[width=8.3cm,angle=0]{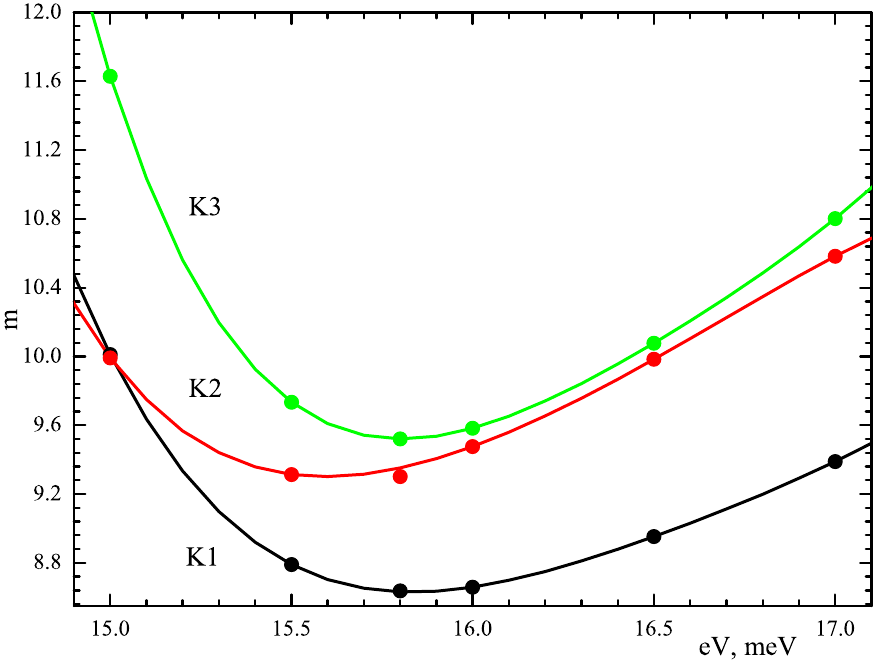}
\caption[]{Dependences of the magnitudes of correction curves on their position on the energy axis for the point-contact presented in Fig.\hyperref[Fig2]{2.1}}.
\label{Fig4}
\end{figure}
\begin{figure}[]
\includegraphics[width=8.3cm,angle=0]{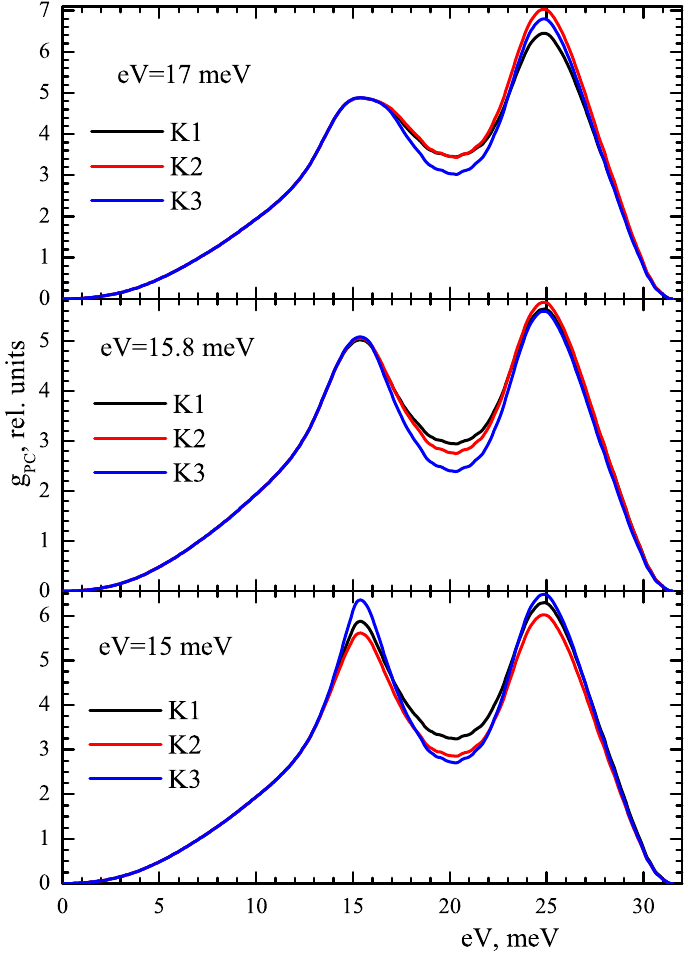}
\caption[]{Variations in the form of the EPI function reconstructed for the contact presented in Fig.\hyperref[Fig2]{2.1}, in dependence of the form of correction curve and its position on the energy axis. The central part of the figure corresponds to the minima of correction curve magnitudes.}
\label{Fig5}
\end{figure}

\begin{figure*}[]
\includegraphics[width=16.5cm,angle=0]{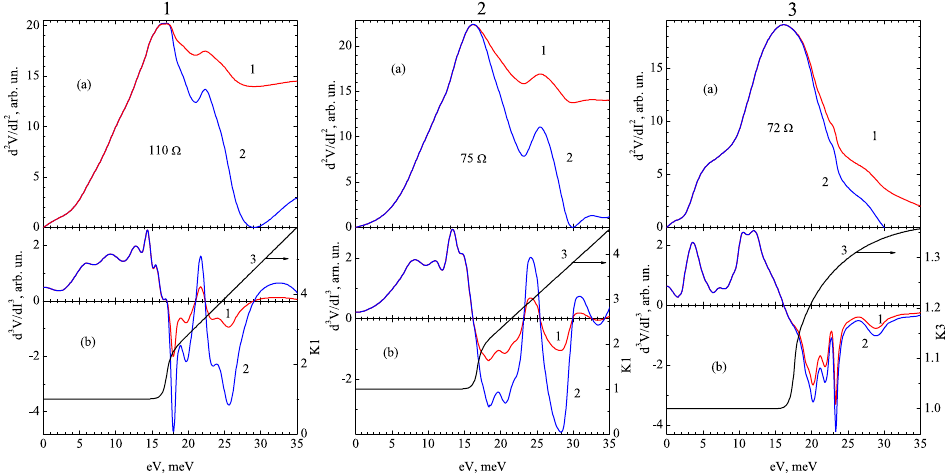}
\caption{Procedure of background correction in the spectra of point-contacts presented in Fig.\hyperref[Fig1]{1}. (a) The second derivatives of $I–V$ curves: (1) initial and (2) after the correction. (b) The third derivatives of the $I–V$ curve, obtained by numerical differentiation of the curves in panels (a) (curves 1); curves after the correction (2); and correction curves $K$ (3).}
\label{Fig6}
\end{figure*}

Since correction curves always have a step-rise portion, it is convenient to simulate their shape using a preform consisting of segments of exponentials. Figure\hyperref[Fig3]{3.1(a)} shows the preforms that were integrated to obtain correction curves of different forms in Fig.\hyperref[Fig3]{3.1(b)}. The initial portions of the curves are identical, whereas the descending portions in Fig.\hyperref[Fig3]{3.1(a)} differ. The initial portion for curve 1 is similar to that for curve 2, reflected and compressed along the $y$ axis; in addition, this portion is extended along the $x$ axis. For curve 3, the initial portion is a combination of two segments of exponentials, matched by a polynomial. When carrying out calculations, the position of the correction curve was identified with the position of the maximum in the preform.

All three correction curves solved successfully the above-stated correction problem in a fairly wide range of their positions on the energy axis for the point-contacts under study. As an example, Fig.\hyperref[Fig3]{3.2(a)} shows a change in the amplitude of the correction curve K1 (see the contact in Fig.\hyperref[Fig2]{2.1} in dependence of its position on the energy axis, as well as the form of all three correction curves at the same displacement, approximately at the center of the interval (Fig.\hyperref[Fig3]{3.2(b)}). To describe quantitatively the correction curve, independent of its shape, it is convenient to use its magnitude, i.e., the effective area under the curve from zero to the boundary of the phonon spectrum, normalized by the value Vmax, corresponding to the spectrum boundary:

\begin{equation}
\label{eq__1}
{m=\frac{1}{V_{max}} \int\limits_{0}^{V_{max}}{\left[K(\omega)-1\right]d\omega}}
\end{equation}

Figure\hyperref[Fig4]{4} shows the dependences of correction curve magnitudes on displacement, and Fig.\hyperref[Fig5]{5} demonstrates variations in the shape of EPI function for different correction curves and their positions on the energy axis for the same point contact. As can be seen, the smallest spread in the shape of EPI function corresponds to the minima of correction curve magnitudes.

\section{BACKGROUND CORRECTION IN POINT-CONTACT YANSON SPECTRA}

Almost all second derivatives of $I–V$ curves of point-contacts at displacements above the phonon spectrum boundary do not turn to zero, whereas the EPI functions at these displacements are identically zero. According to the existing theories, the background is due to the reabsorption of nonequilibrium phonons by electrons in the point-contact \cite{23,24,25}. A quantitative characteristic of the background level $\gamma$ in Yanson spectra is the ratio of the magnitude of the second derivative of $I–V$ curve at the phonon spectrum boundary to its maximum value. However, even under the assumption that nonequilibrium phonons are completely blocked in the contact, this theory cannot explain the existence of spectra with the back­ground level $\gamma\geq0.5$ \cite{26}. To date, the existence of spectra with a large background level has not been physically justified; therefore, some empirical self­consistent iterative procedures are generally applied in practice, which satisfy the requirements of coincidence (within experimental error) for EPI functions of different contacts made of identical materials \cite{3}. This is generally valid for the background level $\gamma\leq0.3$. For the spectra with a large background level, the final result of its subtraction will be determined by the procedure chosen.

\begin{figure}[]
\includegraphics[width=8.5cm,angle=0]{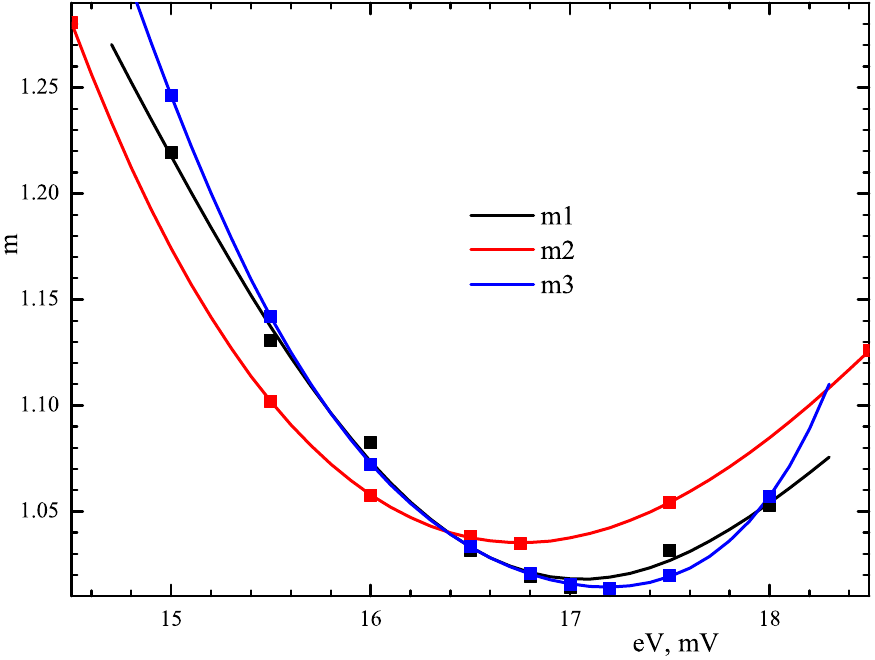}
\caption[]{Dependences of the magnitudes of correction curves on their position on the energy axis for the point-contact shown in Fig.\hyperref[Fig6]{6.1}.}
\label{Fig7}
\end{figure}
\begin{figure}[]
\includegraphics[width=8.5cm,angle=0]{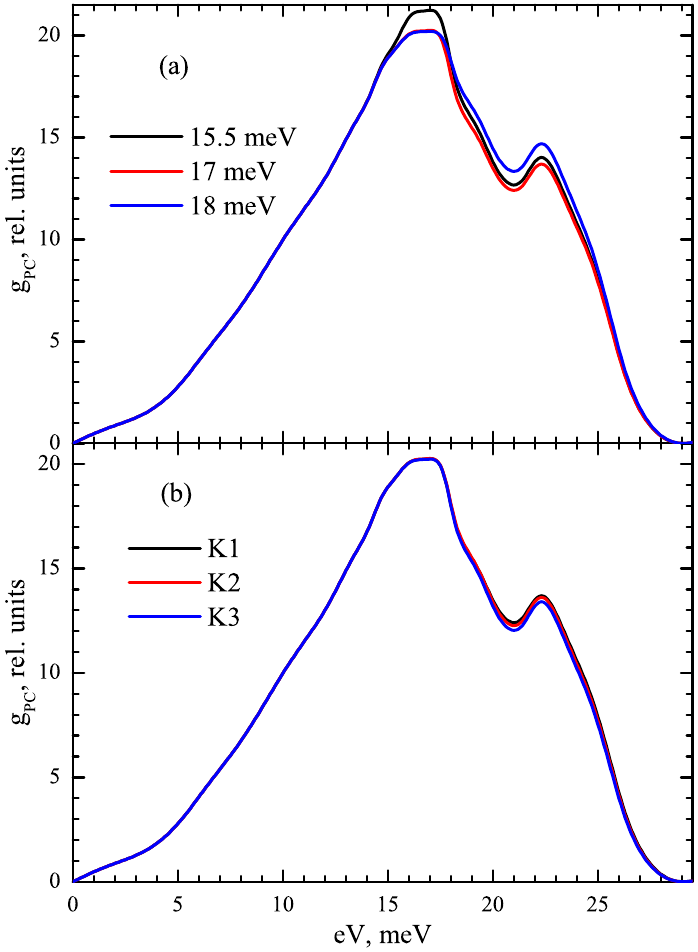}
\caption[]{Variations in the forms of the EPI function reconstructed for the contact presented in Fig.\hyperref[Fig6]{6.1} (a) in dependence of the position of correction curve $K1$ on the energy axis and (b) for different correction curves at the points of minima of their magnitudes.}
\label{Fig8}
\end{figure}
\begin{figure}[]
\includegraphics[width=8.3cm,angle=0]{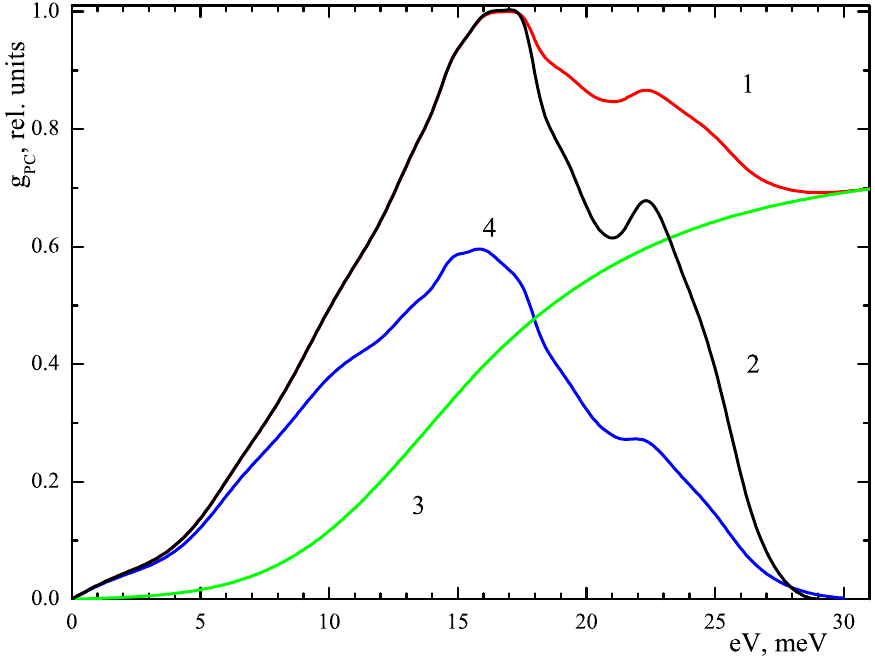}
\caption[]{Comparison of the procedures of background subtraction and correction: (1) initial experimental curve with a background, (2) curve after the correction, (3) traditionally used background curve, and (4) curve after background subtraction. The EPI parameter $\lambda$ for curve 2 is larger by a factor of 1.5 than that for curve (4), and the maximum of curve (4) is about 0.6 of the maximum of curve (2).}
\label{Fig9}
\end{figure}
\begin{figure}[]
\includegraphics[width=8.5cm,angle=0]{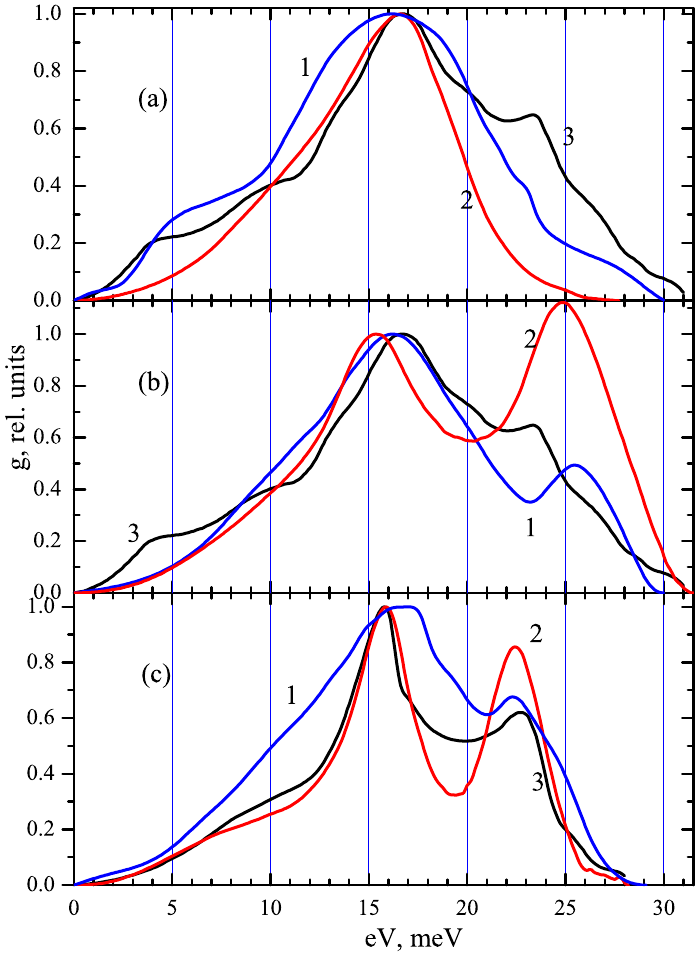}
\caption[]{Comparison of the EPI functions reconstructed from the spectra of contacts in the normal and superconducting states and from the tunnel data. All curves are normalized to unity in the vicinity of the first peak. (a) (1) Amorphous material in the normal state (Fig.\hyperref[Fig6]{6.3}), (2) amorphous material in the superconducting state (Fig.\hyperref[Fig2]{2.3}), and (3) reconstruction from the tunnel data for amorphous films \cite{15}. (b) (1) Defect material in the normal state (Fig.\hyperref[Fig6]{6.2}), (2) defect material in the superconducting state (Fig.\hyperref[Fig2]{2.1}), and (3) reconstruction from the tunnel data for amorphous films \cite{15}. (c) (1) Distorted material in the normal state (Fig.\hyperref[Fig6]{6.1}), (2) defect-free material in the superconducting state (Fig.\hyperref[Fig6]{2.2}), and (3) reconstruction from the tunnel data for crystalline films \cite{27}.}
\label{Fig10}
\end{figure}

At the same time, having compared the curves in Figs.\hyperref[Fig1]{1} and \hyperref[Fig2]{2с}, one can suggest by analogy that the background in Yanson spectra may arise as a result of the suppression of high-frequency phonons, which should manifest itself in violation of the sum rule already for the third derivative of $I–V$ curve. The suppression of high-frequency phonons and, correspondingly, occurrence of background are caused by impurities and lattice defects. Phonons reflect the vibrational structure of material in their generation region. For low-energy phonons with a large wavelength, all these local lattice distortions are averaged on the wave­length and do not affect much the generation conditions. At the same time, generation of high-energy phonons (having a wavelength comparable with the lattice period) near lattice distortions will be hindered, because they are diverse and randomly located. Thus, it follows from this suggestion that a correlation should be observed between the degree of suppression of high-frequency phonons and the background level, which corresponds in principle to the experiment \cite{12,14}.

Derivatives of $I–V$ curves are generally recorded using the standard modulation technique, which is generally accepted in the point-contact and tunnel spectroscopy. Since the modulation voltage makes spectra diffuse, it is taken as low as possible, in order to keep a reasonable compromise between the resolution and noise level. Therefore, the use of the initial experimental curve describing the second derivative of $I–V$ curve to obtain the \emph{third derivative} by numerical differentiation leads in most cases to unacceptably high noise level in the curve obtained. The experimental curves were smoothed to suppress noise: they were initially divided into overlapping segments, which were then approximated by polynomials. The initial experimental points and results of their polynomial approximation are presented in Fig.\hyperref[Fig1]{1}.

Since the procedure of background correction in Yanson spectra by reconstructing the sum rule on the third derivative of $I–V$ curve in no way differs technically from the similar procedure performed on the second derivative of $I–V$ curve of superconducting contacts, we applied the same algorithm.

Figure\hyperref[Fig6]{6} shows the results of application of this algorithm to the contacts presented in Fig.\hyperref[Fig1]{1}. The dependences of the magnitudes of correction curves on their position on the energy axis (Fig.\hyperref[Fig7]{7}) and the variations in the shape of derivatives after the back­ground correction (Fig.\hyperref[Fig8]{8}) are given for different types of correction curves and their position on the energy axis for one of the contacts. As can be seen in Fig.\hyperref[Fig8]{8}, the smallest variations in the form of EPI functions correspond to the minima of correction curve magnitudes. Here, the much smaller magnitude of correction curves (in comparison with the superconducting spectra; compare Figs.\hyperref[Fig4]{4} and \hyperref[Fig7]{7}) leads to almost complete identity of curves after the background correction; i.e., the final result is independent of the form of correction curve.

For Yanson spectra, which are characterized by a high background level, its subtraction by conventional methods \cite{3} leads, first, to a significant distortion of the reconstructed EPI function and, second, to underestimated values of the EPI function $g_{PC}$ and the EPI parameter $\lambda$ (Fig.\hyperref[Fig9]{9}). For the spectrum presented in Fig.\hyperref[Fig9]{9}, the parameter $\lambda$ for the back­ground correction is larger by a factor of 1.5 than for the background subtraction, and the value of the EPI function at the point of maximum for background subtraction is about 60\% of that in the case of correction. Thus, the use of the background correction method makes it possible to refine the form of EPI functions and numerical parameters for the materials whose spectra are difficult to obtain with a low back­ground level.

It is of interest to compare the results of reconstructing EPI functions from the spectra of point-contacts in the superconducting and normal states, as well as to consider them in comparison with the tunnel data \cite{15,27} (Fig.\hyperref[Fig10]{10}).

As follows from the comparison, the obtained EPI functions can arbitrarily be separated into three groups. The EPI function in Fig.\hyperref[Fig10]{10(a)} corresponds to amorphous niobium. Proceeding from the diffusion of curves, the largest degree of amorphization is observed for function 2, reconstructed from a superconducting spectrum, whereas significant recrystallization occurs for the tunnel EPI function (curve 3). In Fig.\hyperref[Fig10]{10(b)}, the degrees of recrystallization are similar, the second peak is blue-shifted in all curves, and their strong broadening is observed. Finally, Fig.\hyperref[Fig10]{10(c)} presents the curves corresponding to the most perfect lattice. The peak positions coincide for all curves, and their diffusion is also much lower than in the previous cases. At the same time, the EPI function reconstructed from the spectrum in the normal state is significantly broadened, and the second peak is suppressed; the second peak in the EPI function reconstructed from tunnel measurements is significantly suppressed. Curve 2, reconstructed from the superconducting spectrum, appears to be most appropriate.

\section{RESULTS AND DISCUSSION}

The existing theories of the spectroscopy of EPI in superconducting point-contacts with direct conductivity consider the ballistic flight of carriers through the constriction and are valid for only superconductors with a large coherence length. The diffusion limit is considered for only a long channel with dirty edges, whereas no theoretical model has been developed yet for inhomogeneous point-contacts. Nevertheless, studies \cite{5,6} form a solid theoretical foundation, the quintessence of which is the proof that the EPI function can be reconstructed from the excess­current nonlinearity related to the scattering of Andreev electrons from nonequilibrium phonons. In the experimental verification of the results of \cite{5,6} it was found that the superconducting background (which was not predicted by the theory) must be taken into account to reconstruct the EPI function. Of much greater importance is the fact that the range of application of the theory was found to be wider than the authors suggested initially: the reconstruction of the EPI function from excess-current nonlinearity turned out to be possible not only for ballistic contacts made of superconductors with a very large coherence length but also for other superconductors with a smaller coherence length, as well as for inhomogeneous contacts. By analogy with the conventional point-contact spectroscopy, on the history scale, we are in the interval by which the Kulik–Omel’yanchuk–Shekhter (KOSh) theory had been developed \cite{1} but the theory explaining the observation of EPI spectra in diffusion point-contacts \cite{11} had not been developed yet. The theory explaining the occurrence of background in Yanson spectra was also absent then. Incidentally, the theoretical explanation of the possibility of occurrence of background at the level above 50\% is still absent. Thus, the experimental results presented here may yield useful information for further development of the theory.

Point-contact spectroscopy is based on carrier duplication: in the current-conducting state electrons are divided into two groups, for which the energy difference between the occupied and free electron states on the Fermi surface is $eV$, i.e., corresponds to the applied voltage \cite{1,2}. If inhomogeneous contacts in the superconducting state provide non-dissipative transmission of electrons through the defect region (i.e., electron duplication is preserved), the second derivative of the $I–V$ curves of these structures is similar to the difference of the $I–V$ curves of the normal and superconducting states of ballistic point-contacts. With allowance for the experimental results for niobium point-contacts, even partially dissipative mode of transmission through the defect region may be applicable for reconstructing the EPI function. On the one hand, this circumstance opens additional possibilities, increasing the number of objects making it possible to gain in some cases information of better quality about the EPI from an unperturbed material in the vicinity of generation of nonequilibrium phonons, as was demonstrated above. At the same time, one must take into account possible presence of a series resistance. In the case of a superconductor with a known position of phonon features, this requirement is easy to satisfy, but a material with unknown phonon spectrum calls for measuring a fairly large number of contacts with different resistances.

Both the superconducting spectra of inhomogeneous contacts and Yanson spectra demonstrate suppression of high-energy phonons, although the physical reasons for this suppression are fundamentally different.

In inhomogeneous superconducting contacts, a nonlinearity is formed due to the scattering of non­equilibrium phonons from the electrons undergoing Andreev reflection, in a volume with a characteristic size on the order of reduced coherence length. The probability of this scattering depends strongly on the concentration of both components, and there are some threshold concentrations of both Andreev electrons and nonequilibrium phonons, below which the probability of their interaction sharply decreases. These threshold concentrations are interrelated: an increase in the concentration of one component decreases the threshold concentration of the other. The concentration of Andreev electrons is determined by the excess current, while the concentration of non­equilibrium phonons is controlled by the density of phonon states. With an increase in the voltage across the contact, the value excess current gradually decreases in the entire energy range. In regards to the density of phonon states, it first increases and, having reached the energy of the first phonon peak, starts rapidly decreasing, thus causing a sharp drop of scattering probability. Another extremely important factor is the dependence of the effective scattering cross section of nonequilibrium phonons from Andreev electrons on the energy of these phonons. Nonequilibrium phonons are not point objects, and the scattering efficiency decreases with an increase in their energy.

Thus, specifically the combination of all these factors leads to the suppression of high-energy phonons. As was shown in the study of superconducting tantalum point-contacts \cite{5}, the degree of this suppression depends very strongly on the rate with which the excess current decreases, whereas the suppression onset is always related to the energy corresponding to the sharp decrease in the density of phonon states. For the contacts considered here, the minima of correction curve magnitudes lie in the energy ranges of $15.8\div16$~meV for the contacts with an ordered lattice (with allowance for the voltage correction for the $50~\Omega$ contact) and $16.5\div16.8$~meV for the amorphous contact.

The suppression of high-energy phonons in Yanson spectra is due to the lattice distortions (induced by impurities and defects). A number of conditions must be satisfied to allow an electron with an excess energy corresponding to the applied voltage to loose this energy via emission of a nonequilibrium phonon. First, the energy of this phonon should coincide with the excess electron energy, i.e., the vibrational lattice parameters in the vicinity of the phonon generation point should allow for this process. Second, the phonon should be able to propagate throughout the crystal; i.e., the vibrational lattice parameters in neighboring volumes should not differ significantly. When the voltage applied to the contact is low, the emitted phonons have a large wavelength, and the defects do not affect much the possibility of phonon generation, because many atoms can be placed within a distance equal to the wavelength, and the lattice parameters are averaged. At the same time, if the phonon wavelength is comparable with lattice parameters, its distortions may hinder phonon generation. Indeed, there will be individual vibrational lattice parameters in the vicinity of each defect, which will differ from those of un perturbed lattice and, possibly, other defects. In addition, random arrangement of defects hinders the propagation of the phonons generated in their vicinity. Thus, the effective generation volume for short-wavelength phonons is smaller than that for long-wavelength phonons, due to which their intensity decreases and general spectral broadening occurs.

The minima of correction curve magnitudes in Yanson spectra are located at energies of $16.6\div17.2$~meV for contacts with an ordered lattice and at $17.6$~meV for amorphous contacts. Thus, the minima of correction curve magnitudes in Yanson spectra are blue-shifted by approximately 1~meV in comparison with the spectra of the superconducting contacts. This pattern correlates with the observed increase in the degree of suppression of high-frequency phonons with a decrease in their wavelength.

So different physical reasons responsible for the suppression of high-energy phonons in the spectra of inhomogeneous superconducting contacts and the spectra of Yanson point-contacts obviously lead to different results when reconstructing the form of EPI functions. The sum-rule reconstruction on the third derivatives of I–V curves of Yanson spectra removes the background on the second derivatives but, however, does not eliminate the reason for its occur­rence. Therefore, the form of the EPI function corresponds to a perturbed material, which is especially pronounced for spectra with a large background and, correspondingly, large magnitude of correction curves: high-frequency peaks are suppressed, and phonon features are broadened. Since the nonlinearity formation domain is removed from the surface (where the lattice defect concentration is maximum) in inhomogeneous superconducting point-contacts, the probability of obtaining spectra corresponding to unperturbed material is higher for these contacts than for Yanson contacts. This remark is especially true for materials with a problematic surface, which are technologically complex for point-contact spectroscopy.

Despite the fact that the derivation of EPI functions from the spectra of inhomogeneous superconducting contacts involves double transformation (subtraction of superconducting background and subsequent correction of the high-energy part of the curves to make the latter satisfy the sum rule), the variance of the obtained results is not of crucial importance. The superconducting background corresponds at large displacements to the general shape of the spectrum, and its small variation barely leads to any significant changes in the form and position of phonon features on the reconstructed EPI function. A change in the form of the correction curve also affects little the form of the reconstructed EPI function if its position on the energy axis corresponds to the magnitude minimum. Nevertheless, there is a strong relationship between the variance in the form of reconstructed EPI function and the correction curve magnitude. There are some small variations in form for curves in Fig. 5 at the points of minima of correction-curve magnitudes (m = 8.6–9.5), whereas for the contact with a resis-tance of 110 . (see Fig. 1) with a background level . ~0.7 the variations in the forms of EPI function after the reconstruction (see Fig. 8, m ~ 1) are negligible.

A possible factor hindering curve processing is the occurrence of nonspectral features (related to the destruction of superconductivity in the near-contact region) in derivatives of I–V curves. These features may be of different nature, for example, thermal or related to nonequilibrium processes. They are not reproducible, and their position depends on temperature and point-contact resistance. Therefore, a rather large volume of experimental data must be gained to exclude these features from consideration.

\section{BRIEF CONCLUSIONS}
\begin{enumerate}
\item Inhomogeneous microcontacts exhibit a radical difference between the second derivatives of $I–V$ curves in the normal and superconducting states: the phonon-related features on the derivatives are either absent or significantly weakened in the normal state. Specific features arise on the second derivatives of $I–V$ curves in the superconducting state, which make it possible to reconstruct the EPI function.
\item A necessary and sufficient condition for reconstructing the EPI function with the aid of inhomogeneous superconducting contacts is to use the second derivative of $I–V$ curve in the superconducting state.
\item The background in the second derivatives of $I–V$ curves of point-contacts in the normal state is caused by lattice defects, which reduce the contribution of high-energy phonons to the spectrum and, correspondingly, violate the sum rule even in the third derivatives of $I–V$ curve. Background correction is performed by reconstructing the broken sum rule on this derivative.
\end{enumerate}

 This study was supported by the National Academy of Sciences of Ukraine, project FTs 4-19.

\end{document}